\documentclass[a4paper,12pt]{article}
\pdfoutput=1
\usepackage{geometry}
\usepackage{amsmath,amssymb,amsfonts}
\usepackage{xcolor,graphicx,cite,soul}
\usepackage{array,booktabs,longtable}
\usepackage{caption,microtype}
\usepackage{subcaption}
\usepackage{hyperref}
\usepackage{datetime}
\usepackage{appendix}

\newcommand{\gsim}
{\;\raisebox{-.3em}{$\stackrel{\displaystyle >}{\sim}$}\;}

\newcommand\al{\alpha}
\newcommand\be{\beta}
\newcommand\tb{\tan\beta}

\newcommand\CBA{c_{\beta - \alpha}}

\newcommand\ReDiag{\mathop{%
  \raise .5pt\hbox{[}%
  \widetilde{\mathrm{Re}}%
  \raise .5pt\hbox{]}}}
\newcommand\ReOffDiag{\mathop{%
  \raise .5pt\hbox{$\llbracket$}%
  \widetilde{\mathrm{Re}}%
  \raise .5pt\hbox{$\rrbracket$}}}

\newcommand\Mh{m_h}
\newcommand\MH{m_H}
\newcommand\MA{m_A}
\newcommand\MHp{m_{H^\pm}}

\newcommand\msq{m_{12}^{2}}

\newcommand\refeq[1]{Eq.~(\ref{#1})}

\newcommand\refta[1]{Tab.~\ref{#1}}
\newcommand\refse[1]{Sect.~\ref{#1}}

\newcommand\citere[1]{Ref.~\cite{#1}}
\newcommand\citeres[1]{Refs.~\cite{#1}}

\newcommand{\CP}{{\cal CP}}
\newcommand{\cp}{{\CP}}

\newcommand{\tev}{\,\, \mathrm{TeV}}
\newcommand{\gev}{\,\, \mathrm{GeV}}

\newcommand\HB{\texttt{HiggsBounds}}
\newcommand\HS{\texttt{HiggsSignals}}

\newcommand\fb{\ensuremath{\mbox{fb}}}

\newcommand{\br}{\text{BR}}
\newcommand{\De}{\Delta}

\newcommand{\sig}{\sigma}

\def\reffi#1{\mbox{Fig.~\ref{#1}}}

\def\la{\lambda}
\newcommand\kala{\ensuremath{\kappa_{\lambda}}}
\newcommand\laSM{\ensuremath{\lambda_{\mathrm{SM}}}}
\newcommand{\lahhh}{\ensuremath{\la_{hhh}}}
\newcommand{\lahhH}{\ensuremath{\la_{hhH}}}
\newcommand{\lahHH}{\ensuremath{\la_{hHH}}}
\newcommand{\lahAA}{\ensuremath{\la_{hAA}}}
\newcommand{\lahHpHm}{\ensuremath{\la_{hH^+H^-}}}

\newcommand{\inter}[2]{\ensuremath{[#1, #2]}}

\definecolor{Orange}{named}{orange}
\definecolor{Purple}{named}{purple}
\definecolor{Lightblue}{cmyk}{0.9,0.1,0.1,0.3}
\definecolor{dgelborange}{cmyk}{0.,0.3,0.5, 0.}
\definecolor{Lila}{rgb}{0.5,0.,1}
\definecolor{Darkgreen}{rgb}{0.,.7,0.2}

\graphicspath{{figs/}}
\renewcommand{\textminus}{$-$}

\allowdisplaybreaks
\begin{document}
\thispagestyle{empty}

\def\thefootnote{\fnsymbol{footnote}}

%\begin{flushleft}
%\mbox{Compiled on \today, at \currenttime}
%\end{flushleft}
\begin{flushright}
\mbox{}
IFT--UAM/CSIC-21-054\\
\end{flushright}

\vspace{0.5cm}

\begin{center}

{\large\sc 
  {\bf Sizable triple Higgs couplings in the 2HDM:\\[.3em]
    Prospects for future $e^+e^-$ colliders}}%
\footnote{Talk presented at the International Workshop on Future
  Linear Colliders (LCWS2021),\\
  \mbox{}\hspace{5mm} 15-18 March 2021. C21-03-15.1} 

\vspace{1cm}

{\sc
F.~Arco$^{1,2}$%
\footnote{Speaker}%
\footnote{email: Francisco.Arco@uam.es}% 
, S.~Heinemeyer$^{2,3,4}$%
\footnote{email: Sven.Heinemeyer@cern.ch}%
~and M.J.~Herrero$^{1,2}$%
\footnote{email: Maria.Herrero@uam.es}%
%\footnote{former address}%
}

\vspace*{.7cm}

{\sl
$^1$Departamento de F\'isica Te\'orica, 
Universidad Aut\'onoma de Madrid, \\ 
Cantoblanco, 28049, Madrid, Spain

\vspace*{0.1cm}

$^2$Instituto de F\'isica Te\'orica (UAM/CSIC), 
Universidad Aut\'onoma de Madrid, \\ 
Cantoblanco, 28049, Madrid, Spain

\vspace*{0.1cm}

$^3$Campus of International Excellence UAM+CSIC, 
Cantoblanco, 28049, Madrid, Spain 

\vspace*{0.1cm}

$^4$Instituto de F\'isica de Cantabria (CSIC-UC), 
39005, Santander, Spain

}

\end{center}

\vspace*{0.1cm}

\begin{abstract}
\noindent

In the framework of the $\mathcal{CP}$ conserving Two Higgs Doublet
Model (2HDM), type~I and~II, we study the triple Higgs couplings with
at least one light $h$ Higgs boson that is identified by the 125 GeV
Higgs boson. We define benchmark planes that exhibit large values of
triple Higgs couplings, while being in agreement with all experimental
and theoretical constraints. Finally, we analyze the impact of the
triple Higgs couplings on the production cross
section of two neutral Higgs bosons in two channels,
$\sig(e^+e^-\to h_i h_j Z)$ and $\sig(e^+e^- \to h_i h_j \nu\bar{\nu})$
with $h_i h_j = hh, hH, HH, AA$.
We show that the triple Higgs couplings have an important impact on
these $e^+e^-$ production cross sections.
\end{abstract}

%\pacs{}

\def\thefootnote{\arabic{footnote}}
\setcounter{page}{0}
\setcounter{footnote}{0}

\newpage

%%%%%%%%%%%%%%%%%%%%%%%%%%%%%%%%%%%%%%%%%%%%%%%%%%%%%%%%%%%%%%%%%%%%%%%%%%%%%%%
%%%%%%%%%%%%%%%%%%%%%%%%%%%%%%%%%%%%%%%%%%%%%%%%%%%%%%%%%%%%%%%%%%%%%%%%%%%%%%%

\section{Introduction}
\label{sec:intro}

The discovery of a Higgs boson at ATLAS and CMS in 2012 was a
milestone in high-energy physics~\cite{Aad:2012tfa,Chatrchyan:2012xdj}.
Within theoretical and experimental uncertainties this new particle is
consistent with the existence of a Standard-Model~(SM) Higgs boson at a mass
of~$\sim 125 \gev$~\cite{Khachatryan:2016vau}.
The measurements of Higgs-boson couplings, which are known
experimentally to a precision of roughly $\sim 20\%$, leave room for
Beyond Standard-Model (BSM) interpretations. Many BSM models possess
extended Higgs-boson sectors. While no sign of BSM physics was (yet)
discovered at the LHC, one of the main tasks of the
LHC Run~III and beyond is to determine whether 
the particle forms part of the Higgs sector of an extended
model.

The measurement of the trilinear Higgs coupling of the SM-like Higgs
boson \lahhh\ is a key element in the investigation of the Higgs-boson sector.
(see \citeres{deBlas:2019rxi, DiMicco:2019ngk} for recent reviews on
Higgs couplings measurements at future colliders). In
the case of a BSM Higgs-boson sector, equally important is the
measurement of BSM 
trilinear Higgs-boson couplings. So far, most experimental studies
have assumed the SM value of \lahhh. In BSM models, however,
this coupling may differ significantly
from its SM value. The expected precision that can be achieved at
different future colliders in the measurement of \lahhh\ can depend
strongly on the value realized in nature. 

A natural extension of the Higgs-boson sector of the SM is the ``Two Higgs
Doublet Model'' (2HDM) (see,
e.g.,~\cite{Gunion:1989we,Aoki:2009ha,Branco:2011iw} for reviews). The 2HDM
contains five physical Higgs bosons: the 
light and the heavy $\CP$-even $h$ and $H$, the $\CP$-odd $A$, and the pair of
charged Higgs bosons, $H^\pm$.
The two mixing angles $\al$ and $\be$ diagonalize the $\CP$-even and $\CP$-odd
Higgs boson sector, respectively.
The parameter $\tb$ is defined by the ratio of the two vacuum
expectation values, $\tb := v_2/v_1$.

We review the allowed ranges for all triple Higgs couplings involving at
least one light, SM-like Higgs boson. Specifically: \lahhh, 
\lahhH, \lahHH, \lahAA\ and \lahHpHm~\cite{Arco:2020ucn}.
It is assumed that the
light $\CP$-even Higgs-boson $h$ is SM-like with a mass of 
$\Mh \sim 125 \gev$. All other Higgs bosons are assumed to be heavier.
To avoid flavor changing neutral currents at the
tree-level, a $Z_2$~symmetry is imposed~\cite{Glashow:1976nt}.
This symmetry is possibly softly broken by the parameter $\msq$.
Four types of the 2HDM can be realized, 
depending on how this symmetry is extended to the fermion sector:
type~I and~II, lepton specific and flipped~\cite{Aoki:2009ha}.
In the 2HDM also the stability conditions for the Higgs
potential change with respect to the SM~\cite{Deshpande:1977rw} (for a
review see~\cite{Bhattacharyya:2015nca}).   
Here (as in \citere{Arco:2020ucn}) we focus on the 2HDM type~I and~II. 
The allowed ranges for the triple Higgs couplings are obtained taking
into account: the theoretical 
constraints from unitarity and stability (we use
\cite{Bhattacharyya:2015nca,Akeroyd:2000wc,Barroso:2013awa}),
the experimental production and decay rates of the SM-like Higgs
boson (we use \HS~\cite{Bechtle:2013xfa,Bechtle:2014ewa,Bechtle:2020uwn}),
experimental constraints from direct Higgs-boson searches (we use
\HB~\cite{Bechtle:2008jh,Bechtle:2011sb,Bechtle:2013wla,Bechtle:2015pma,Bechtle:2020pkv}),
as well as flavor observables (we use
\texttt{SuperIso}~\cite{Mahmoudi:2008tp,Mahmoudi:2009zz},
complemented with~\cite{Li:2014fea,Cheng:2015yfu,Arnan:2017lxi})  
and electroweak precision observables (EWPO) (we use~the $S$, $T$
and~$U$ parameters~\cite{Peskin:1990zt,Peskin:1991sw}, complemented with
\cite{Grimus:2007if,Funk:2011ad} and bounds from
\cite{Tanabashi:2018oca}).  

The main interest in the ranges allowed for the triple Higgs-boson
couplings is the fact that 
they affect the rates of multi-Higgs boson production at
current and future colliders.
Within the 2HDM, the production  of Higgs boson pairs like
$hh$, $hH$, $HH$, $hH^\pm$, $AA$ and $H^+H^-$ can be
significantly affected by (large) values of triple Higgs couplings,
which are yet allowed by the present constraints. 
At $e^+e^-$ colliders two different channels are of interest:
$e^+e^-\to h_ih_jZ$ and $e^+e^-\to h_ih_j\nu\bar{\nu}$
with $h_ih_j=hh$, $hH$, $HH$ and $AA$.
The first one is similar to the “Higgs-strahlung” channel of single
Higgs production. The second one has an important contribution
from the vector-boson fusion mediated subprocess, $W^+W^- \to h_ih_j$,
where the $WW$ pairs (virtual, in general) are radiated from the
initial $e^+e^-$ together with the neutrinos:
$e^+e^- \to W^\ast W^\ast\nu\bar{\nu}$. 
The latter processes also receives a contribution from the $Z$
mediated subprocess, $e^+e^- \to Zh_ih_j$, with $Z \to\nu\bar{\nu}$,
which is usually smaller than the contribution from $WW$~fusion at the
high energy colliders. 

%%%%%%%%%%%%%%%%%%%%%%%%%%%%%%%%%%%%%%%%%%%%%%%%%%%%%%%%%%%%%%%%%%%%%%%%%%%%%%
%%%%%%%%%%%%%%%%%%%%%%%%%%%%%%%%%%%%%%%%%%%%%%%%%%%%%%%%%%%%%%%%%%%%%%%%%%%%%%

\section{The Model and the constraints}

\subsection{The Two Higgs Doublet Model}
\label{sec:2hdm}

We work within the $\cp$ conserving 2HDM. The scalar potential of this model
is given by~\cite{Branco:2011iw}:
\begin{eqnarray}
V &=& m_{11}^2 (\Phi_1^\dagger\Phi_1) + m_{22}^2 (\Phi_2^\dagger\Phi_2) - \msq (\Phi_1^\dagger
\Phi_2 + \Phi_2^\dagger\Phi_1) + \frac{\la_1}{2} (\Phi_1^\dagger \Phi_1)^2 +
\frac{\la_2}{2} (\Phi_2^\dagger \Phi_2)^2 \nonumber \\
&& + \la_3
(\Phi_1^\dagger \Phi_1) (\Phi_2^\dagger \Phi_2) + \la_4
(\Phi_1^\dagger \Phi_2) (\Phi_2^\dagger \Phi_1) + \frac{\la_5}{2}
[(\Phi_1^\dagger \Phi_2)^2 +(\Phi_2^\dagger \Phi_1)^2]  \;,
\label{eq:scalarpot}
\end{eqnarray}
\noindent
where $\Phi_1$ and $\Phi_2$ denote the two $SU(2)_L$ doublets.
As mentioned above, the occurrence of tree-level flavor
changing neutral currents (FCNC) is avoided by imposing a $Z_2$ symmetry 
on the scalar potential. 
The scalar fields transform as:
\begin{align}
  \Phi_1 \to \Phi_1\;, \quad \Phi_2 \to - \Phi_2\;.
  \label{eq:2HDMZ2}
\end{align}
The $Z_2$ symmetry, however, is softly broken by the $\msq$ term in
the Lagrangian. The extension of the $Z_2$ symmetry to the Yukawa
sector avoids tree-level FCNCs. 
Depending on the $Z_2$ parities of the fermions, this results in four
variants of 2HDM: type I, type II, lepton-specific (or type X) and 
flipped (or type Y)~\cite{Aoki:2009ha}.

Taking the electroweak symmetry breaking (EWSB) minima to be
neutral and $\cp$-conserving, the scalar fields after EWSB
can be parameterized as:%
\begin{eqnarray}
\Phi_1 = \left( \begin{array}{c} \phi_1^+ \\ \frac{1}{\sqrt{2}} (v_1 +
    \rho_1 + i \eta_1) \end{array} \right) \;, \quad
\Phi_2 = \left( \begin{array}{c} \phi_2^+ \\ \frac{1}{\sqrt{2}} (v_2 +
    \rho_2 + i \eta_2) \end{array} \right) \;.
\label{eq:2hdmvevs}
\end{eqnarray}
Here $v_1, v_2$ are the real vevs acquired by the fields
$\Phi_1, \Phi_2$, respectively, and $\tb := v_2/v_1$. They satisfy the 
relation $v = \sqrt{(v_1^2 +v_2^2)}$, with $v \simeq 246\gev$ being the SM vev.
There are eight degrees of freedom in the fields $\phi_{1,2}^\pm$,
$\rho_{1,2}$ and $\eta_{1,2}$. They give rise to three Goldstone
bosons, $G^\pm$ and $G^0$, 
and five massive physical scalar fields: two $\cp$-even scalar fields,
$h$ and $H$, one $\cp$-odd one, $A$, and one charged pair, $H^\pm$.
Here the mixing angles $\alpha$ and $\beta$ diagonalize the CP-even and
-odd Higgs boson sectors, respectively.

After the consideration of the minimization conditions, the 2HDM
can be characterized by seven free parameter. The 2HDM is studied
in the physical basis, where the free parameters
of the model, which we use as input, are chosen as:
\begin{equation}
c_{\be-\al} \; , \quad \tb \;, \quad v \; ,
\quad \Mh\;, \quad \MH \;, \quad \MA \;, \quad \MHp \;, \quad \msq \;,
\label{eq:inputs}
\end{equation}
where we use the short-hand notation
$s_x = \sin(x)$, $c_x = \cos(x)$.
The analysis identifies the lightest $\cp$-even Higgs boson,
$h$, with the one observed at $\sim 125 \gev$.

The potential of the 2HDM defines the interactions in the scalar
sector. Here we focus on the couplings of the
lightest $\cp$-even Higgs boson with the other BSM bosons: \lahhh, 
\lahhH, \lahHH, \lahAA\  and \lahHpHm.
These $\la_{h h_i h_j}$ couplings are defined
such that the Feynman rules are given by:
\begin{equation}
	\begin{gathered}
		\includegraphics{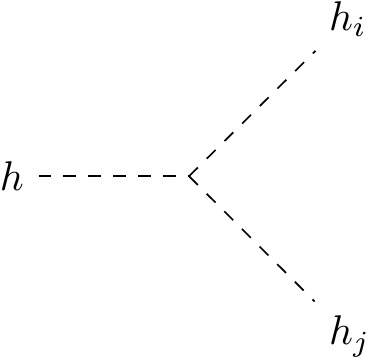}
	\end{gathered}
	=- i\, v\, n!\; \la_{h h_i h_j}
\label{eq:lambda}
\end{equation}
Here $n$ is the number of identical particles in the vertex, 
and we adopt this notation so the light Higgs
trilinear has the same definition as in the SM, i.e. $-6iv\laSM$ with
$\laSM=\Mh^2/2v^2\simeq0.13$. 

It should be noted that all the couplings of the $\cp$-even Higgs bosons
strongly depend on $\CBA$. For $\CBA=0$ one recovers all the
interactions of the SM Higgs boson for the $h$ state, which is called
the \textit{alignment limit}. This limit is very important since
the Higgs measurements in
colliders seem to overall agree with the SM values, see the discussion
in the next subsection.

%%%%%%%%%%%%%%%%%%%%%%%%%%%%%%%%%%%%%%%%%%%%%%%%%%%%%%%%%%%%%%%%%%%%%%%%%%%%%%

\subsection{Constraints}
\label{sec:const}

Here we briefly list the experimental and theoretical constraints that
were employed in the analysis (more details can be found in
\citere{Arco:2020ucn}).

\begin{itemize}
\item {\bf Constraints from electroweak precision data}\\
Constraints from the electroweak precision observables (EWPO)
can for ``pure'' Higgs-sector extensions of the SM, 
be expressed in terms of the oblique parameters $S$, $T$ and
$U$~\cite{Peskin:1990zt,Peskin:1991sw}.
In the 2HDM the $T$~parameter is most
constraining and requires either $\MHp \approx \MA$ or $\MHp \approx \MH$.
In \citere{Arco:2020ucn} we explored three scenarios:
(A)\;$\MHp = \MA$, (B)\;$\MHp = \MH$ and (C)\;$\MHp = \MA = \MH$.
Here we will focus on scenario~C.
For the evaluation of the $T$ parameter we use the code
\texttt{2HDMC-1.8.0}~\cite{Eriksson:2009ws}.

\item {\bf Theoretical constraints}\\
Here the important constraints come from
tree-level perturbartive unitarity and stability of the vacuum.
They are ensured by an explicit test on the underlying Lagrangian
parameters, see \citere{Arco:2020ucn} for details.
The parameter space allowed by these two constraints can be enlarged, in
particular to higher BSM Higgs-boson mass values by the condition,
\begin{equation}
  \msq = \frac{\MH^2\cos^2\al}{\tb}~.
  \label{eq:m12special}
\end{equation}

\item {\bf Constraints from direct searches at colliders}\\
The $95\%$ confidence level
exclusion limits of all important searches for BSM Higgs bosons
are included in the public code
\HB\,\texttt{v.5.9}~\cite{Bechtle:2008jh,Bechtle:2011sb,Bechtle:2013wla,Bechtle:2015pma,Bechtle:2020pkv},
including Run~2 data from the LHC.
Given a set of theoretical
predictions in a particular model, \HB\ determines which is the most
sensitive channel and determines, based on this most sensitive
channel, whether the point is allowed or not at the $95\%$~CL.
As input the code requires some specific predictions from the model,
like branching ratios or Higgs couplings, that we computed with the
help of \texttt{2HDMC}~\cite{Eriksson:2009ws}.

\item {\bf Constraints from the SM-like Higgs-boson properties}\\
Any model beyond the SM has to accommodate the SM-like Higgs boson,
with mass and signal strengths as they were measured at the LHC.
%LHC~\cite{Aad:2012tfa,Chatrchyan:2012xdj,Khachatryan:2016vau}.
In our scans the compatibility of the $\cp$-even scalar $h$ with a mass
of $125.09\gev$ with the measurements of signal strengths at Tevatron and 
LHC
is checked with the code
\texttt{HiggsSignals v.2.6}~\cite{Bechtle:2013xfa,Bechtle:2014ewa,Bechtle:2020uwn}. 
\texttt{HiggsSignals} provides a
statistical $\chi^2$ analysis of the SM-like Higgs-boson predictions of
a certain model compared to the measurement of Higgs-boson signal rates
and masses from Tevatron and LHC. Again, the predictions of the 2HDM
have been obtained with {\tt{2HDMC}}~\cite{Eriksson:2009ws}.
Here, as in \citere{Arco:2020ucn}, we will require that for a  parameter
point of the 2HDM to be allowed, the corresponding $\chi^2$ is within
$2\,\sig$ ($\De\chi^2 = 6.18$)
from the SM fit.

\item {\bf Constraints from flavor physics}\\
Constraints from flavor physics have proven to be very significant
in the 2HDM mainly because of the presence of the charged Higgs boson.
Various flavor observables like rare $B$~decays, 
$B$~meson mixing parameters, $\br(B \to X_s \gamma)$, 
LEP constraints on $Z$ decay partial widths
etc., which are sensitive to charged Higgs boson exchange, provide
effective constraints on the available 
parameter space~\cite{Enomoto:2015wbn,Arbey:2017gmh}. 
Here we take into account the decays $B \to X_s \gamma$ and
$B_s \to \mu^+ \mu^-$, which are most constraining. This is done with
the code \texttt{SuperIso}~\cite{Mahmoudi:2008tp,Mahmoudi:2009zz}
where the model input is given by {\tt{2HDMC}}.  We have modified the
code as to include the Higgs-Penguin type corrections in $B_s \to
\mu^+ \mu^-$ which were not included in the original version of
\texttt{SuperIso}~\cite{Mahmoudi:2008tp,Mahmoudi:2009zz}. These
corrections are indeed relevant for the present work since these
Higgs-Penguin contributions are the ones containing the potential
effects from triple Higgs couplings in $B_s \to \mu^+ \mu^-$.  
\end{itemize}

%%%%%%%%%%%%%%%%%%%%%%%%%%%%%%%%%%%%%%%%%%%%%%%%%%%%%%%%%%%%%%%%%%%%%%%%%%%%%%
%%%%%%%%%%%%%%%%%%%%%%%%%%%%%%%%%%%%%%%%%%%%%%%%%%%%%%%%%%%%%%%%%%%%%%%%%%%%%%

\section{Triple Higgs couplings}

%%%%%%%%%%%%%%%%%%%%%%%%% F I G U R E %%%%%%%%%%%%%%%%%%%%%%%%%%%%%%%%%%%%%%%%%
% FIGURE C4
\begin{figure}[p]
\begin{center}
	%{\small 2HDM type I, scenario C, $\msq = (\MH^2\cos^2\al)/(\tb)$}
	
	\includegraphics[height=0.25\textheight]{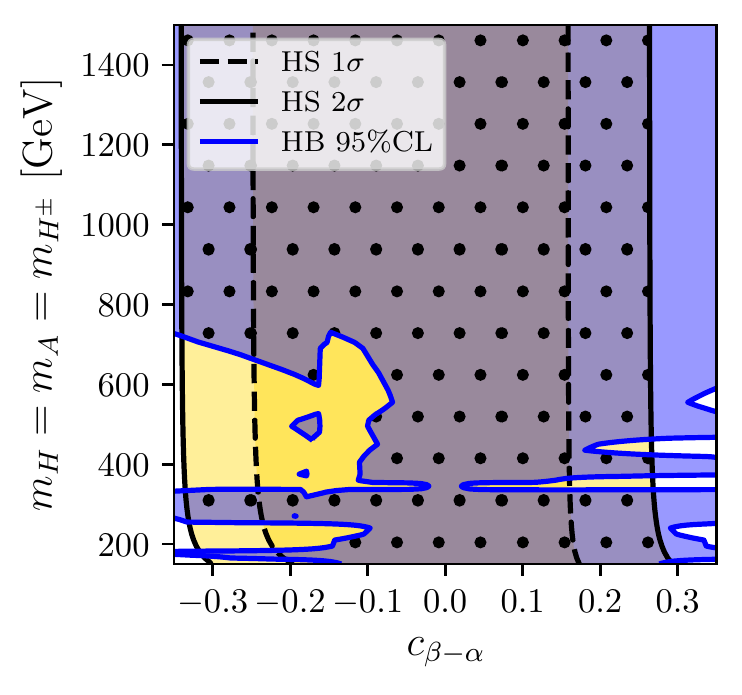}\includegraphics[height=0.25\textheight]{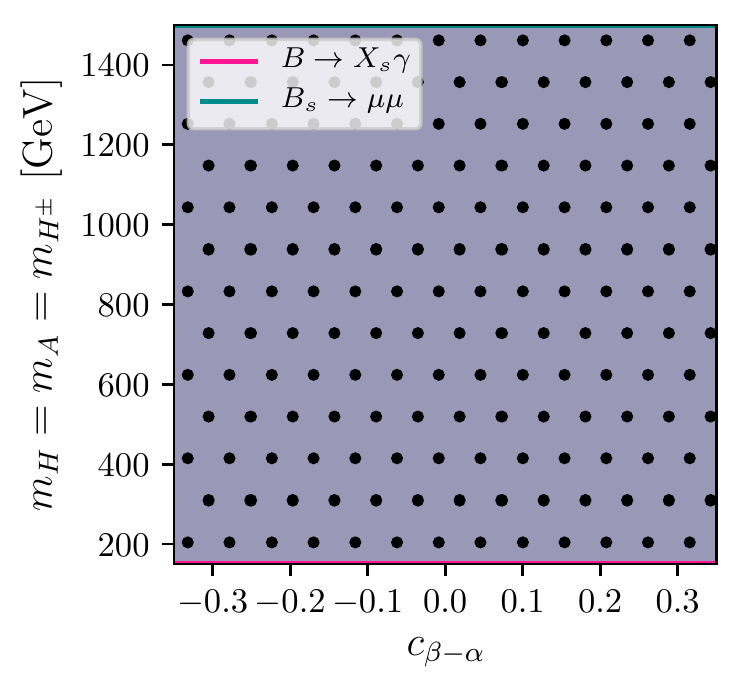}
	\includegraphics[height=0.25\textheight]{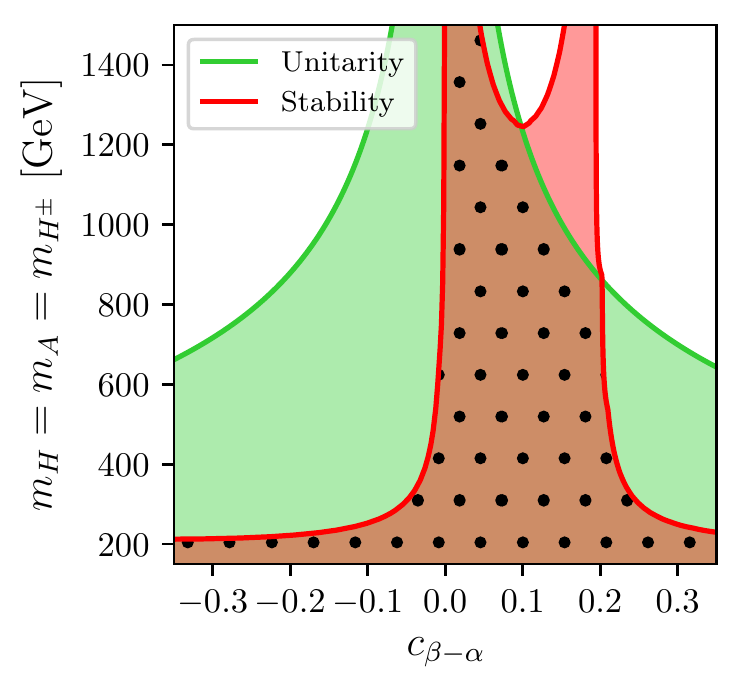}\includegraphics[height=0.25\textheight]{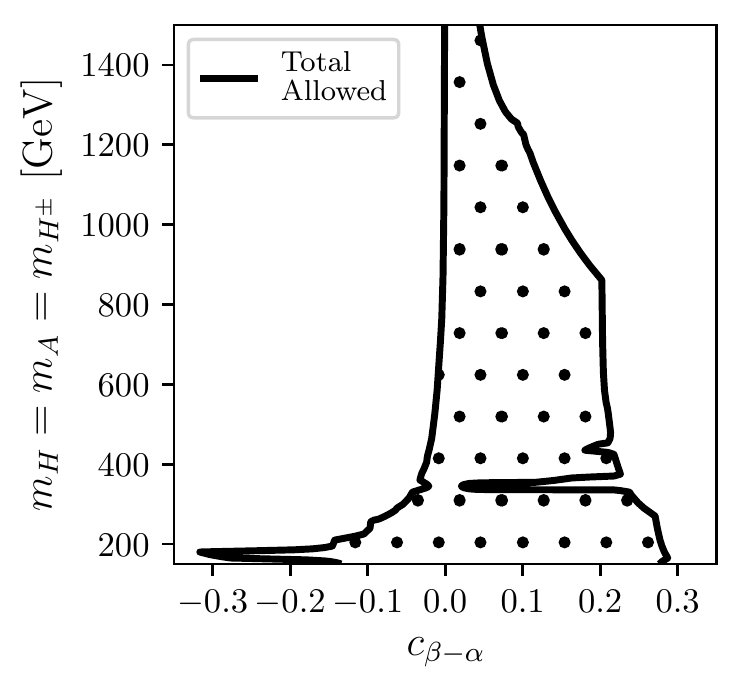}
	\includegraphics[height=0.4\textheight]{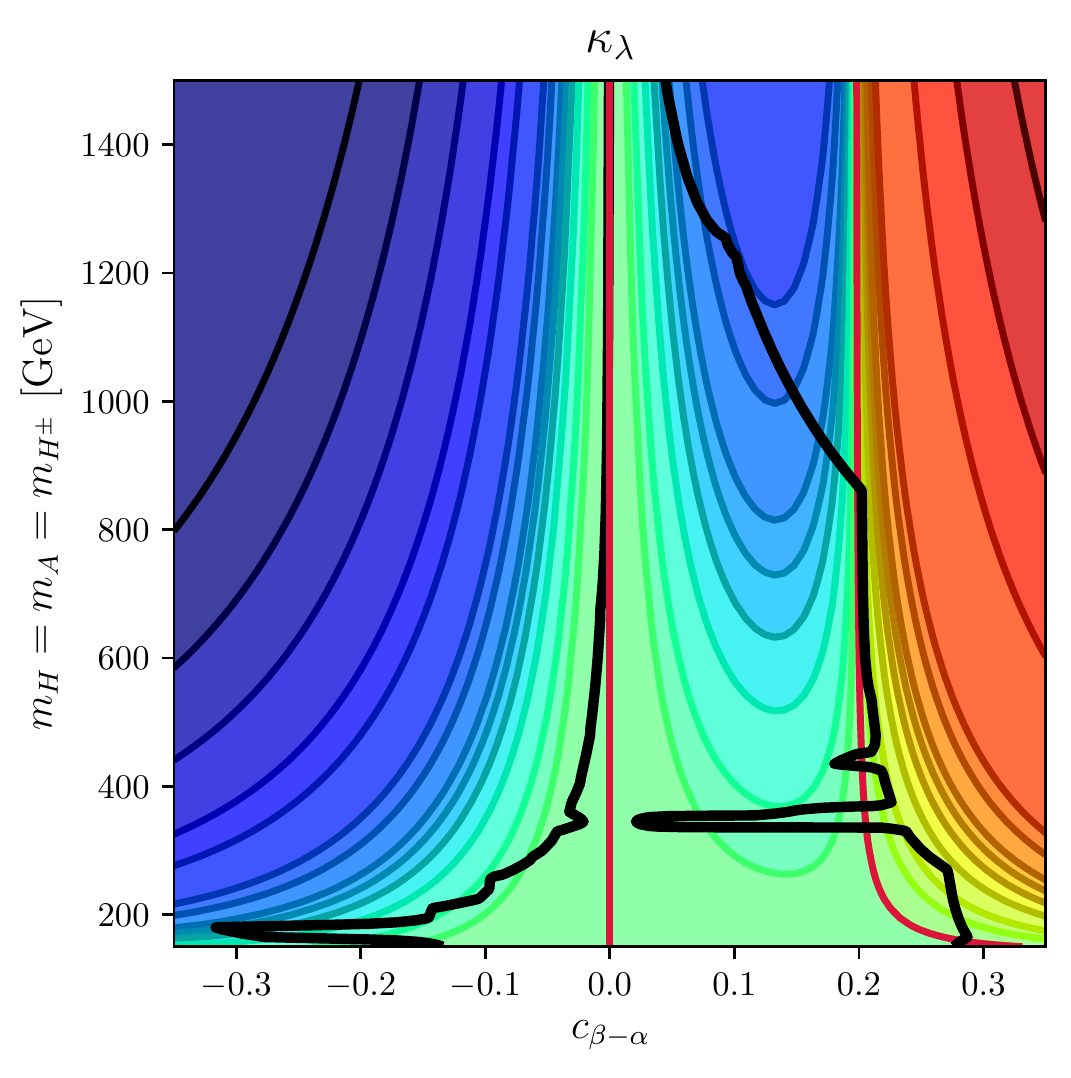}\includegraphics[height=0.4\textheight]{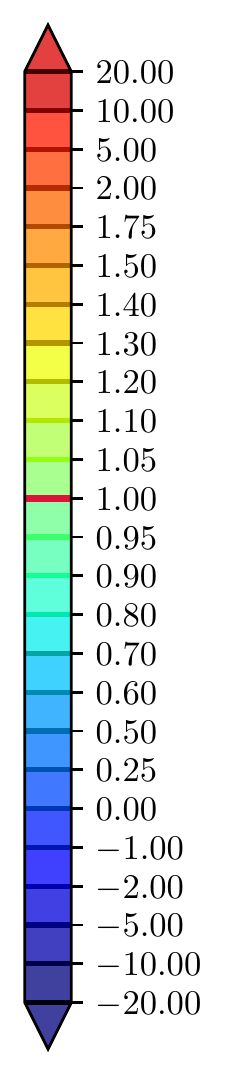}
	\caption{
{\bf (A)} 
Predictions for $\kala = \lahhh/\laSM$ in the  2HDM type~I, scenario~C,
in the $(\CBA, m)$ plane with $m = \MH = \MA = \MHp$,
$\msq = (\MH^2\cos^2\al)/(\tb)$ and $\tb = 10$.
\emph{Upper right plot:} Allowed areas by flavor physics from
$B \to X_s \gamma$ (pink),
$B_s \to \mu^+ \mu-$ (teal) and both (dotted). 
\emph{Middle left plot:} Allowed areas by the theoretical constraints from unitarity (green), stability  (red) and  both (dotted).
\emph{Middle right plot:} Total allowed area (dotted). 
\emph{Lower big plot:} Contour lines of $\kala = \lahhh/\laSM$. 
Red contours correspond to $\kala=1$.
The thick solid contours is the boundary of the total allowed area.} 
\label{fig:C1-cba-MHp}
\end{center}
\end{figure}
%%%%%%%%%%%%%%%%%%%%%%%%% F I G U R E %%%%%%%%%%%%%%%%%%%%%%%%%%%%%%%%%%%%%%%%%
%%%%%%%%%%%%%%%%%%%%%%%%% F I G U R E %%%%%%%%%%%%%%%%%%%%%%%%%%%%%%%%%%%%%%%%%
\begin{figure}[t]\ContinuedFloat
\begin{center}
	\begin{subfigure}[b]{0.7\textwidth}
		\includegraphics[height=0.25\textheight]{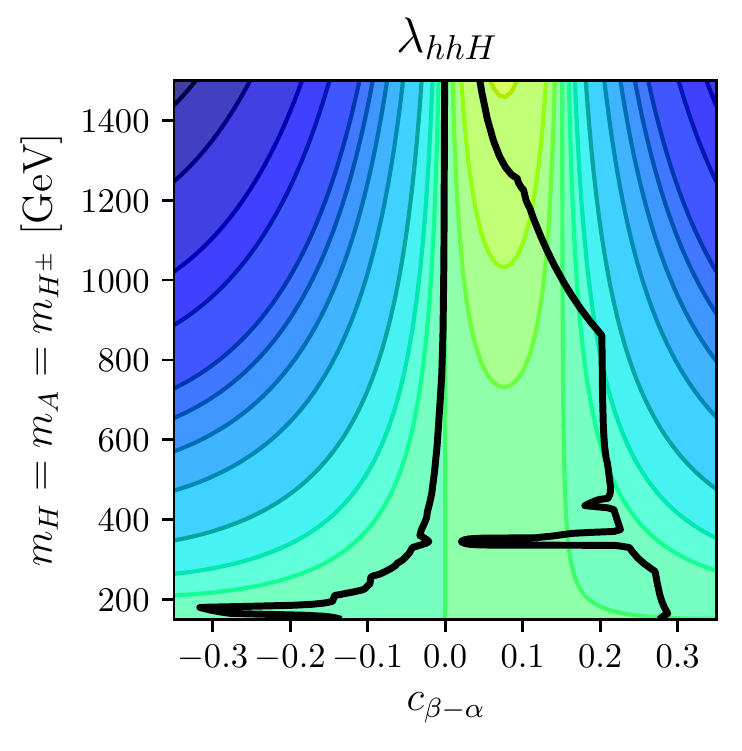}\includegraphics[height=0.25\textheight]{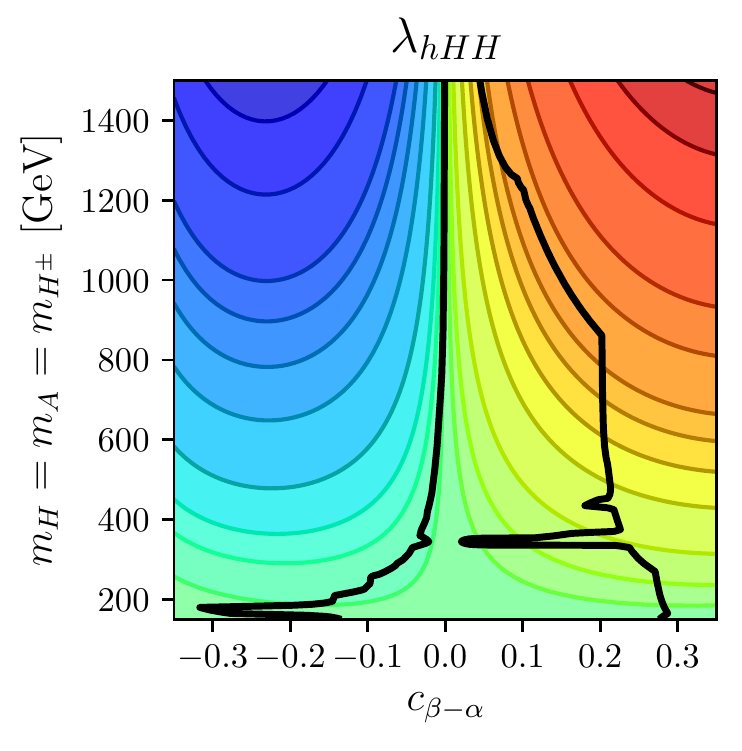}
		\includegraphics[height=0.25\textheight]{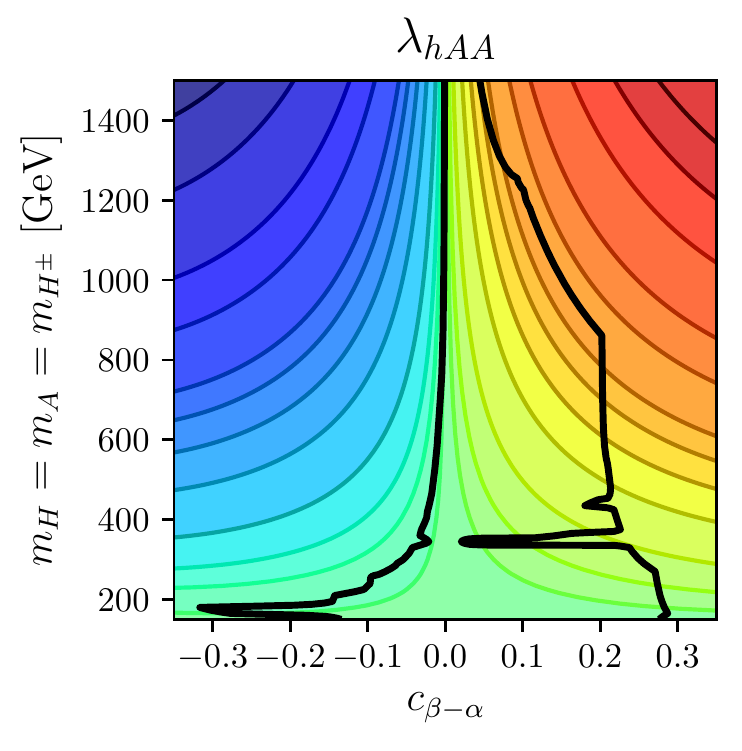}\includegraphics[height=0.25\textheight]{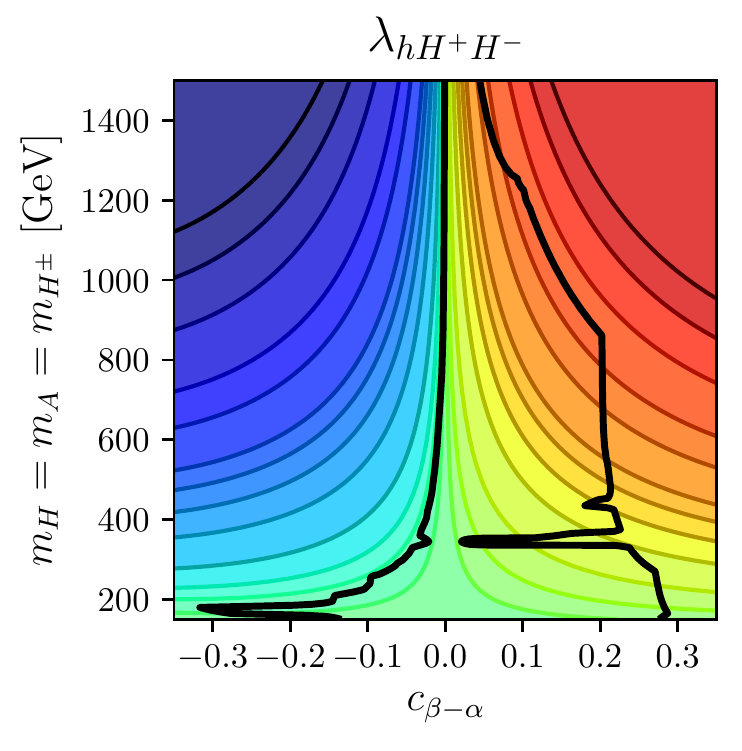}
	\end{subfigure}	
	\begin{subfigure}[b]{0.1\textwidth}
		\includegraphics[height=0.48\textheight]{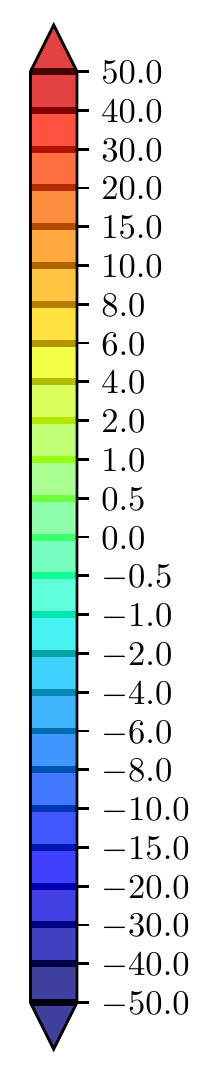}
	\end{subfigure}
\caption{
{\bf (B)} 
Contour lines for triple Higgs couplings in the  2HDM type~I, scenario~C,
in the $(\CBA, m)$ plane for  $m = \MH = \MA = \MHp$,
$\msq = (\MH^2\cos^2\al)/(\tb)$ and $\tb = 10$.
\emph{Upper left:} $\lahhH$,
\emph{upper right:} $\lahHH$,
\emph{lower left:} $\lahAA$,
\emph{lower right:} $\lahHpHm$. 
The thick solid contour is as in \protect\reffi{fig:C1-cba-MHp}(A). 
}
\label{fig:C1-cba-MHp-heavy}
\end{center}
\end{figure}
%%%%%%%%%%%%%%%%%%%%%%%%% F I G U R E %%%%%%%%%%%%%%%%%%%%%%%%%%%%%%%%%%%%%%%%%

In \citere{Arco:2020ucn} several benchmark planes have been defined.
Here we review the results for the ($\CBA$, $\MHp = \MH = \MA \equiv m$)
plane in the 2HDM type~I, with $\tb = 10$ and $\msq$ fixed via
\refeq{eq:m12special}. 
The allowed regions by the constraints considered are displayed in
upper part of \reffi{fig:C1-cba-MHp}(A).
The upper left panel shows the allowed
regions by the collider searches and the direct measurements  
of the light $125 \gev$ Higgs boson, given by \HB\ and \HS\ respectively. 
The allowed region given by \HS\ is $\sim\inter{-0.32}{0.28}$, nearly
independent on $m$. The upper right panel shows the flavor constraints,
that are not relevant for this choice of the input parameters.
In the middle left plot the constraints from unitarity and stability
are displayed, which are the most constraining ones. Positive and
negative values for $\CBA$ are allowed for $m\lesssim 250\gev$ 
but above that mass, due to the stability conditions, only positive
values for $\CBA$ up to $\CBA\sim0.2$ are allowed. For masses larger
than $\sim800\gev$ both stability and unitarity conditions  
narrow the allowed region to the alignment limit.
The middle right panel shows the final allowed region by all the
constrains discussed above, 
that is mainly dominated by the theoretical constraints. The
corresponding (black) outline is shown in all other plots.

The predictions for the triple Higgs couplings with at least one $h$
in this benchmark plane are displayed in the lower panel of
\reffi{fig:C1-cba-MHp}(A) for $\kala$ and in \reffi{fig:C1-cba-MHp}(B)  
in the case of $\lahhH$ (upper left), $\lahHH$ (upper left), $\lahAA$
(lower left) and $\lahHpHm$ (lower right). 
 The values that can be reached by $\kala$ range from $\kala \sim 0.07$
for $\CBA \sim 0.1$ and large $\MHp$ close to $1200 \gev$ to about $\kala
\sim 1.2$ for the largest allowed $\CBA$ values and $\MHp \sim 300 \gev$. 
The values of $\lahhH$ range between $\sim -1.0$ and $\sim1.6$, where that 
maximum value of $\lahhH$ is found for $\CBA\sim0.05$ and large $\MHp$.
Both $\kala$ and $\lahhH$ tend to 1 and 0 respectively in the alignment limit.
 
The ranges reached by the triple Higgs couplings involving two
heavy Higgs boson, as shown in \reffi{fig:C1-cba-MHp-heavy}(B), are found
to be
$\lahHH \sim \inter{-0.2}{12}$, 
$\lahAA \sim \inter{-0.2}{12}$ and
$\lahHpHm \sim \inter{-0.5}{24}$.
For all of them the maximum values are found on the edge for larger $\CBA$ and 
$800\gev \lesssim \MHp \lesssim 1300 \gev$.

In \cite{Arco:2020ucn} we performed a thorough exploration of the five
triple Higgs couplings that involves at least one $h$ state, and the
final allowed ranges of those couplings are gathered in
\refta{tab:THCranges}. The allowed range for $\kala$ is smaller in
type~II compared to type~I, because the collider measurements in
type~II are more stringent and require $\CBA$ to be closer to the
alignment limit than for type~I. For the other non-SM couplings the
ranges are very similar in both type~I and type~II. Here it should be
noted that the maximum values of $\lahHH$, $\lahAA$ and $\lahHpHm$ are
realized in scenarios 
with mass splitting among the heavy Higgs bosons, i.e.\ in the cases
(A) and (B), see \refse{sec:const}. More details can be found in
\citere{Arco:2020ucn}. 
In particular, the large allowed ranges of $\kala$ emphasize the
importance of an evaluation of the sensitivity of future experiments
($e^+e^-$ and $pp$) to $\lahhh$ away from its SM value.

%%%%%%%%%%%%%%%%%%%%%%%%% T A B L E %%%%%%%%%%%%%%%%%%%%%%%%%%%%%%%%%%%%
\begin{table}
\begin{center}
\begin{tabular}{c||c|c}
 & type I & type II\tabularnewline
\hline \hline
$\kala$ & {[}\textminus 0.5, 1.5{]} & {[}0.0, 1.0{]}\tabularnewline
\hline 
$\lahhH$ & {[}\textminus 1.4, 1.5{]} & {[}\textminus 1.6, 1.8{]}\tabularnewline
\hline 
$\lahHH$ & {[}0, 15{]} & {[}0, 15{]}\tabularnewline
\hline 
$\lahAA$ & {[}0,16{]} & {[}0,16{]}\tabularnewline
\hline 
$\lahHpHm$ & {[}0,32{]} & {[}0,32{]}\tabularnewline
\end{tabular}
\caption{Final allowed ranges of $\kala$, $\lahhH$, $\lahHH$, $\lahAA$ and $\lahHpHm$ by all the considered constraints in the 2HDM type I and type II.}
\label{tab:THCranges}
\end{center}
\end{table}
%%%%%%%%%%%%%%%%%%%%%%%%% T A B L E %%%%%%%%%%%%%%%%%%%%%%%%%%%%%%%%%%%%

%%%%%%%%%%%%%%%%%%%%%%%%%%%%%%%%%%%%%%%%%%%%%%%%%%%%%%%%%%%%%%%%%%%%%%%%%%%%%%
%%%%%%%%%%%%%%%%%%%%%%%%%%%%%%%%%%%%%%%%%%%%%%%%%%%%%%%%%%%%%%%%%%%%%%%%%%%%%%

\section{Sensitivity at future $e^+e^-$ colliders}

In this section we discuss the impact of the studied
triple Higgs couplings on the di-Higgs production at future linear
$e^+e^-$ colliders. 
We will consider two different channels: $e^+e^-\to h_ih_jZ$ 
and $e^+e^-\to h_ih_j\nu\bar{\nu}$ with $h_ih_j=hh,hH,HH\ \mathrm{and}\ AA$. 
The cross sections presented in this section where computed with 
$\mathtt{MadGraph}$\cite{Alwall:2014hca} and
$\mathtt{FeynRules}$\cite{Alloul:2013bka}, 
neglecting the electron mass. The width of the Higgs bosons in the
2HDM was computed with $\mathtt{2HDMC}$~\cite{Eriksson:2009ws}.

%%%%%%%%%%%%%%%%%%%%%%%%% FIGURE %%%%%%%%%%%%%%%%%%%%%%%%%%%%%%%%%%%%%%%%%%%%%%
\begin{figure}[htb!]
\begin{center}
	\includegraphics[width=0.49\textwidth]{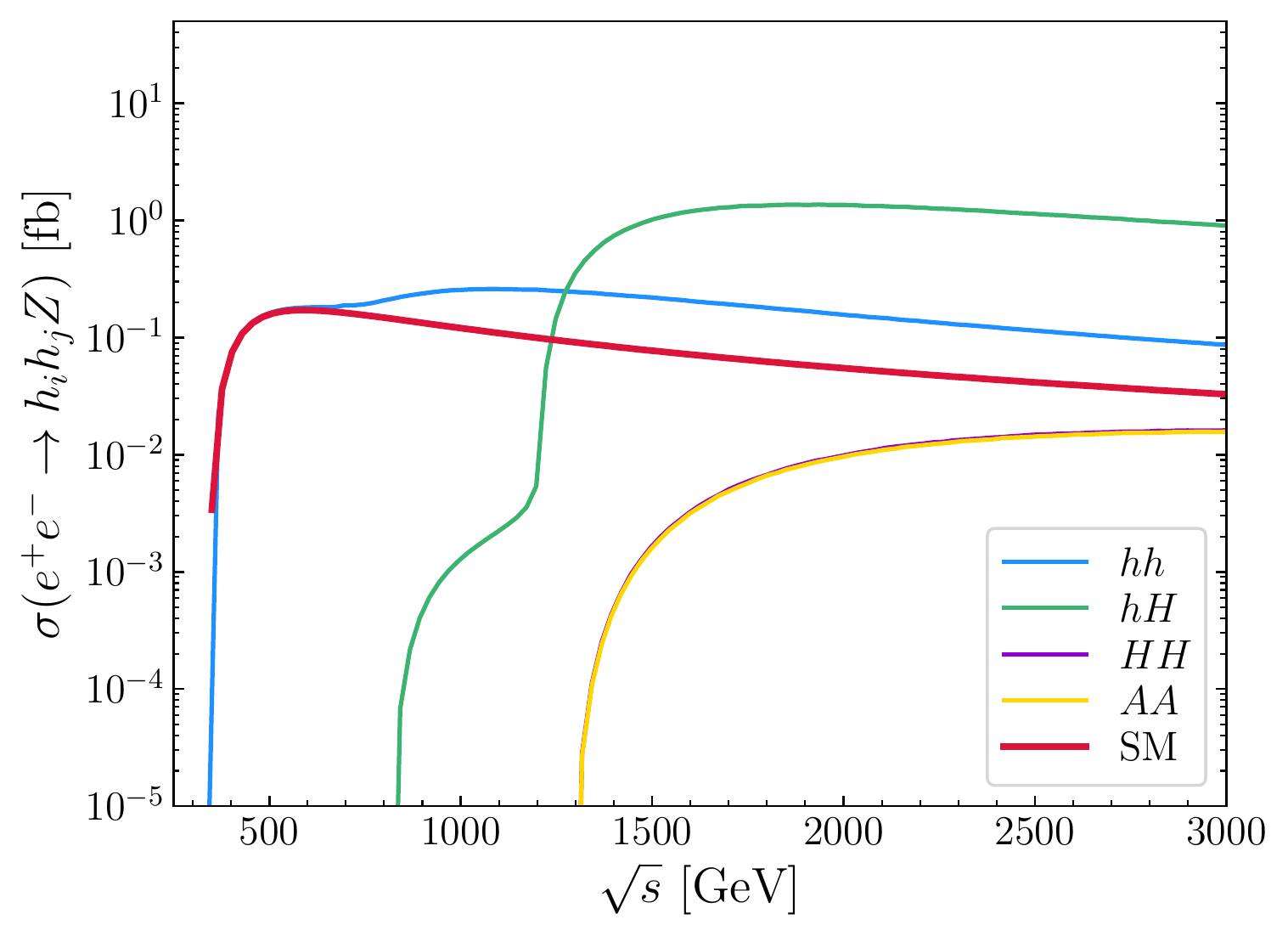}
	\includegraphics[width=0.49\textwidth]{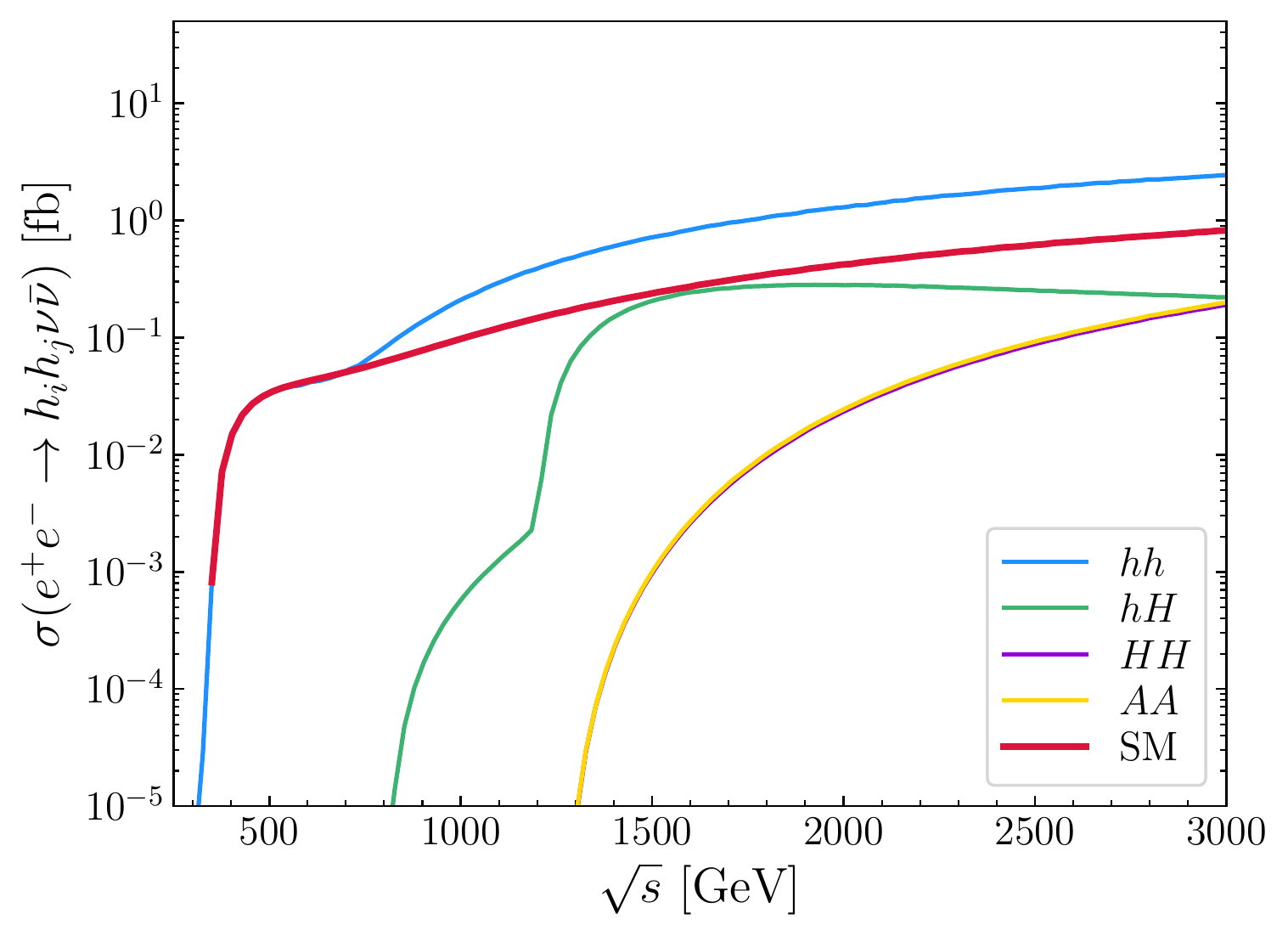}	
\end{center}
\caption{Cross sections as a function of the center-of-mass energy
  $\sqrt{s}$ for the processes $e^+e^-\to h_ih_jZ$ (left) and
  $e^+e^-\to h_ih_j\nu\bar{\nu}$ (right) for the particular point of
  the 2HDM type~I defined by the input parameter values:
  $m_H=m_A=m_{H^\pm}=600$ GeV,  $\tan \beta= 10$, $c_{\beta-\alpha}=
  0.2$,  and $\msq=m_H^2\cos^2\alpha/\tb$ (see text).} 
\label{fig:XSsqrts}
\end{figure}
%%%%%%%%%%%%%%%%%%%%%%%%% FIGURE %%%%%%%%%%%%%%%%%%%%%%%%%%%%%%%%%%%%%%%%%%%%%%

We present in \reffi{fig:XSsqrts} an illustrative example of the
results for the cross sections 
$e^+e^-\to h_ih_jZ$ (left) and $e^+e^-\to h_ih_j\nu\bar\nu$ (right)
as a function of the collider energy.
We chose a particular point of the 2HDM type~I contained in the benchmark plane
discussed in the previous section, with input parameters $m=600\gev$, $\tb=10$, 
$\CBA=0.2$ and $m_{12}^2=m_H^2\cos^2\alpha/\tb$. For this point the following 
values for the triple Higgs couplings are realized: $\kala\simeq1$,
$\lahhH=-0.5$, $\lahHH=\lahAA=6$ and $\lahHpHm=12$.
The value of $\kala \simeq 1$ away from the alignment limit is
realized via ``accidental'' cancellations in the dependence of
$\lahhh$ on the underlying parameters.
One of the main differences between both channels is that all the
cross sections in the $h_ih_jZ$ channel decrease with 
the center-of-mass energy $\sqrt{s}$, while in the case of $h_ih_j\nu\bar{\nu}$ 
the cross sections increases with $\sqrt{s}$ (except for
$hH\nu\bar{\nu}$ in the range of $\sqrt{s}$ analyzed).
The reason behind this is that the diagrams that contributes in the
$Z$ production channel are mediated by a $Z^\ast$, which decrease with
the collision energy, but on the other hand in the neutrino channel
there are diagrams mediated by a pair $W^\ast W^\ast$ 
(known as vector boson fusion or VBF topologies) that are known to
increase with the energy (see, e.g., \citere{Roloff:2019crr} and
references therein). This is indeed the reason why for 
$e^+e^-$ colliders with energies at and above the TeV scale the
diagrams with VBF configuration dominates the $h_ih_j\nu\bar{\nu}$ 
production rates, even though the later also contains $Z^\ast$ mediated 
diagrams.
 
The $hh$ production in both channels (blue lines) shows an enhancement
at $\sqrt{s}\gsim700\gev$ with respect to the SM prediction 
(red lines). This difference 
can be explained with the new contributions coming from
 diagrams mediated by BSM Higgs bosons. The relevant BSM diagrams are
 $e^+e^-\to W^\ast W^\ast\nu\bar{\nu}\to H^{(\ast)}\nu\bar{\nu}
 \to hh\nu\bar{\nu}$ and 
 $e^+e^-\to Z^\ast \to H^{(\ast)} Z\to hhZ\ (\to hh\nu\bar{\nu})$, 
 that contain $\lahhH$, as well as 
 $e^+e^-\to Z^\ast \to hA^{(\ast)} \to hhZ\ (\to hh\nu\bar{\nu})$.
 All of them are resonant in the case that the (virtual) $H^\ast$ or
 $A^\ast$ propagates on-shell. 
 The larger contribution of them is the one coming from $H^\ast$
 mediated diagrams,  
 because the diagrams with a propagating $A^\ast$ is proportional to
 the coupling  $g_{hAZ}\propto \CBA$. Since the experimental data
 requires a SM-like $h$~boson, one finds $\CBA \ll 1$, so that these
 are somewhat suppressed despite being resonant.  
 It is important to note that in this case the narrow width approximation
 does not yield a good approximation to the total cross section,
 because the interference of this resonant diagrams with the 
 rest of the contributions is not negligible.
 
In the case of the $hHZ$ production (green line) we can see that the
cross section can reach values around $\sim1\ \fb$, more than one order
of magnitude larger than $e^+e^-\to hhZ$ in the SM. However, the
dominant contribution here is given by the
resonant diagram $e^+e^-\to Z^\ast\to HA^{(\ast)} \to HhZ$, which does
not depend on any triple Higgs coupling. A consequence of this is that 
the cross section receives a large enhancement around the threshold
$\sqrt{s}\sim m_H+m_A=1200\gev$.
This large contribution is also found in the process
$e^+e^-\to hH\nu\bar{\nu}$, that shows the same behavior with $\sqrt{s}$
as $e^+e^-\to hHZ$, but with a smaller cross section.
 
Finally, we comment the production of two heavy BSM bosons (purple lines for 
$HH$ and yellow lines for $AA$). For our choice of $m_H = m_A$, the 
di-$\mathcal{CP}$-odd production is always very similar to the 
di-$\mathcal{CP}$-even production. (Consequently, the purple $HH$ line
is nearly fully covered by the yellow $AA$ curve.)
The $HHZ$ and $AAZ$ channels have a very small cross section,
only reaching $\sim 10^{-2}\ \fb$ at $\sqrt{s} = 3 \tev$.
On the other hand, the $HH/AA$ production in the neutrino channel
reaches sizable cross section, $\sim0.2\ \fb$ at $3 \tev$, about the
same level as the $hH$ cross section. The $HH\nu\bar\nu$ and
$AA\nu\bar\nu$ channels are mainly 
dominated by the diagram containing $\lahHH$ or $\lahAA$, respectively.

%%%%%%%%%%%%%%%%%%%%%%%%%%%%%%%%%%%%%%%%%%%%%%%%%%%%%%%%%%%%%%%%%%%%%%%%%%%%%%
%%%%%%%%%%%%%%%%%%%%%%%%%%%%%%%%%%%%%%%%%%%%%%%%%%%%%%%%%%%%%%%%%%%%%%%%%%%%%%

\section{Conclusions and summary}

The measurement of the triple Higgs coupling $\lahhh$
is an important task at future colliders. 
Depending on its relative size to the SM value,
certain collider options result in a higher (or lower) experimental accuracy. 
On the other hand, large values of triple Higgs couplings involving heavy Higgs
bosons can lead to relevant effects on the production cross sections
of BSM Higgs bosons.

Within the framework of Two Higgs Doublet Models (2HDM) type~I and~II we
have reviewd the allowed ranges for all triple Higgs couplings involving at
least one light, SM-like Higgs boson~\cite{Arco:2020ucn}.
All relevant theoretical and experimental constraints are taken into
account. This comprises from the theory side 
unitarity and stability conditions. From the
experimental side we require agreement with measurements of the
SM-like Higgs-boson rate as measured at the LHC as well as with
the direct BSM Higgs-boson searches.
Furthermore agreement with flavor observables and
electroweak precision data (where the $T$~parameter plays the most
important role) was required. We have shown one benchmark plane as
example. Taking 
all analyzed benchmark planes into account~\cite{Arco:2020ucn}, the
final allowed ranges for the triple Higgs couplings are obtained; they
are summarized in \refta{tab:THCranges}.

We have also studied how these triple Higgs couplings can affect
the neutral di-Higgs production at future $e^+e^-$ colliders
in two different production modes:
$e^+e^-\to h_ih_j Z$ and $e^+e^-\to h_ih_j\nu\bar{\nu}$
with $h_ih_j=hh,\ hH,\ HH,\ AA$. In an example point, allowed by
all the relevant constraints (and part of the benchmark plane shown),
we find clear signals of BSM Higgs bosons, arising from
the studied triple Higgs couplings. We find that the production of $hh$ can
be affected by $\lahhH$ via the resonant processes
$e^+e^-\to W^\ast W^\ast\nu\bar{\nu}\to H^{(\ast)}\nu\bar{\nu}\to hh\nu\bar{\nu}$.
The $Z$~channel will be important at lower collider energies, such
like ILC, and will be able to give access to this coupling if the mass
of $H$ is sufficiently low to be produced on-shell (a viable
possiblity in the 2HDM).
On the other hand, the neutrino channel will be dominant 
and will have a larger cross section at higher center-of-mass energies
projected for CLIC. In this 
case, the $hh\nu\bar{\nu}$ channel can give access to $\lahhH$ 
for larger values for $m_H$.
In the presented point the deviation from the SM of the $h$ self-coupling, given
by $\kala:=\lahhh/\lahhh^\mathrm{SM}$, is very small due to an
``accidental'' cancellations in the dependence of $\lahhh$ on the
underlying parameters. As a consequence
the effect of this parameter and the corresponding diagrams are very
similar to the SM case. 
Analyses of the accuracy of the measurement of $\laSM$ have been performed, 
see for instance \citeres{Roloff:2019crr,Gonzalez-Lopez:2020lpd} and references
therein, and a similar accuracy can be expected the case of the 2HDM.
The $hH$ production channels do not show any important dependence 
on any of the triple Higgs couplings. On the other hand, the
production cross sections of $HH\nu\bar{\nu}$ and $AA\nu\bar{\nu})$
(which are very similar), 
are dominated by the size of $\lahHH$ or $\lahAA$, respectively.
Therefore, a measurement of these processes would yield a good
opportunity to measure these couplings.

%\newpage

\end{document}